\documentclass{article}

\usepackage{arxiv}

\usepackage[utf8]{inputenc} 
\usepackage[T1]{fontenc}    
\usepackage{hyperref}       
\usepackage{url}            
\usepackage{booktabs}       
\usepackage{amsfonts}       
\usepackage{nicefrac}       
\usepackage{microtype}      
\usepackage{graphicx}
\usepackage{natbib}
\usepackage{doi}

\title{
\texorpdfstring{
  \raisebox{-0.15em}{\includegraphics[height=1.0em]{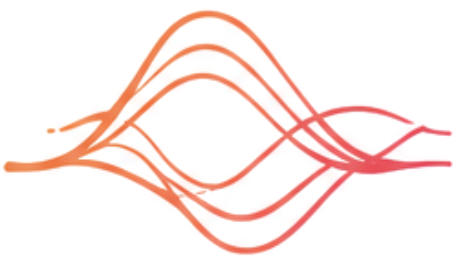}}\hspace{-0.1em}
  Shao: Scaling Acoustic Token Language Models Toward High-Fidelity Music Generation
}{
  Shao: Scaling Acoustic Token Language Models Toward High-Fidelity Music Generation
}
}

\author{
  \large
  \begin{tabular}{c} \textbf{Jiafeng Liu}$^{1,\dagger}$ \\ \texttt{ljiafeng@ccom.edu.cn} \end{tabular}
  \hskip 4em
  \begin{tabular}{c} \textbf{Yuanliang Dong}$^{1,\dagger}$ \\ \texttt{gunterdong@mail.ccom.edu.cn} \end{tabular} \\[5mm]
  \large
  \begin{tabular}{c} {Hongjia Liu$^{1}$ \quad Yuqing Cheng$^{1}$ \quad Zhancheng Guo$^{1}$ \quad Huijing Liang$^{1}$ \quad Wenbo Zhan$^{1}$ } \end{tabular} \\[2mm]
  \large
  \begin{tabular}{c} {Yuming Sun$^{1}$ \quad Xiaobing Li$^{1}$ \quad Feng Yu$^{1}$ \quad Maosong Sun$^{2,*}$} \end{tabular} \\[2mm]
  \normalsize
  \begin{tabular}{c}
    $^{1}$Central Conservatory of Music \quad $^{2}$Tsinghua University
  \end{tabular}
}

\date{}

\renewcommand{\headeright}{}
\renewcommand{\undertitle}{}
\renewcommand{\shorttitle}{Shao: Acoustic Token Language Models For Music Generation}

\hypersetup{
pdftitle={Shao: Scaling Acoustic Token Language Models Toward High-Fidelity Music Generation},
pdfsubject={Music Generation, Audio Language Modeling, Neural Audio Codec},
pdfauthor={Jiafeng Liu, Yuanliang Dong, Hongjia Liu, Yuqing Cheng, Wenbo Zhan, Yuming Sun, Xiaobing Li, Feng Yu, Maosong Sun},
pdfkeywords={music generation, acoustic tokens, audio language model, neural audio codec, residual vector quantization}
}

\begin{document}
\maketitle
\begingroup
\renewcommand{\thefootnote}{}
\footnotetext{$^{\dagger}$Core contributors; equal contribution.}
\footnotetext{$^{*}$Corresponding author.}
\endgroup

\begin{abstract}
	A common design pattern in high-quality music generation is to handle structure and fidelity in different representation spaces: a generator first models high-level structure, followed by diffusion-based or neural decoding stages that reconstruct fine details. In this work, we explore an alternative view: both may be progressively modeled within a single deep acoustic-token hierarchy. To study this, we build a 64-layer residual vector quantization (RVQ) acoustic representation and propose a two-stage coarse-to-fine generation framework. A backbone model first generates coarse acoustic tokens for the full track, and a super-resolution model then completes finer tokens within the same acoustic token space. The super-resolution stage works at full-track scale and refines tokens layer by layer while running in parallel over time, leading to a fixed 62-step inference process. To jointly improve lyric alignment and fine-detail reconstruction, we further introduce hybrid-attention training: the alignment objective uses causal attention, while layer-wise refinement uses full attention. A key finding is that text--vocal alignment can emerge within pure acoustic-token language modeling, without requiring a separate semantic token stage. Moreover, initializing the super-resolution model from the trained backbone significantly improves convergence and final quality. Taken together, our results suggest that high-quality music generation can be effectively pursued without separating structure and fidelity into heterogeneous representation spaces. Instead, both can be progressively modeled within a unified acoustic-token hierarchy, pointing toward a simpler and more unified path to high-quality music generation. Code and model checkpoints are available at \url{https://github.com/Shao-Music-AI/Shao}.
\end{abstract}


\section{Introduction}
	\subsection{Background and Motivation}
	High-quality music generation requires both coherent musical structure and strong acoustic fidelity. The generated music should follow meaningful melody, rhythm, long-range form, and lyric conditioning, while also preserving timbre, transients, and overall perceptual quality. Recent progress has largely been driven by systems that decompose these requirements into separate stages, where high-level structure is modeled first and fine acoustic detail is reconstructed later, often through diffusion-based or neural decoding modules \citep{audiolm,musiclm,stableaudio,diffrhythm,songbloom,acestep15}.
	
	This decomposition has proven effective, but it also leads to a common formulation of music generation: structure and fidelity are often handled in different representation spaces. In this work, we explore a different perspective. Instead of introducing a separate semantic stage or an external acoustic renderer, we investigate whether both can be progressively modeled within a single deep acoustic-token hierarchy.

	\subsection{Key Challenges}

    This direction is challenging for three reasons. First, a pure acoustic-token approach depends heavily on the quality of the underlying audio tokenizer. If the token representation is too shallow, it may preserve coarse content but leave important acoustic detail underrepresented, limiting the fidelity that a downstream language model can ultimately generate. This motivates using a deeper residual quantization hierarchy. However, increasing the number of residual quantization layers also makes the tokenizer harder to train stably, and a deeper hierarchy is only useful if its later layers encode meaningful fine acoustic detail rather than noisy or redundant residuals.
    
    Second, even with such a tokenizer, the resulting acoustic tokens are difficult to model. They are dense, high-rate representations that encode both long-range musical structure and fine acoustic variation. This leads to long sequences with complex dependencies, making naive autoregressive generation hard to train and expensive to decode.
    
    Third, text conditioning introduces an additional modeling challenge. A music generation model must not only generate plausible acoustic content, but also make that content follow the given prompt, duration, and lyrics when available. This becomes especially challenging for vocal music, where the generated audio must remain consistent with the provided lyrics while preserving natural musical expression.
    
    These challenges suggest that a pure acoustic-token approach must solve three problems at once: learning a deep and stable acoustic representation, factorizing generation over many acoustic token layers in a scalable way, and incorporating text conditions effectively into full-length acoustic-token generation.

	\subsection{Overview of Our Approach}
	To address these challenges, we adopt a unified pure acoustic-token paradigm with a 64-layer residual vector quantization (RVQ) hierarchy and a two-stage coarse-to-fine generation framework. A backbone model first generates full-length coarse acoustic tokens, and a super-resolution model then refines higher layers within the same acoustic token space. The super-resolution stage operates at full-track scale and predicts in parallel over time, yielding a fixed 62-step refinement process at inference.
	
	To jointly improve lyric alignment and fine-detail reconstruction, we introduce hybrid-attention training for super-resolution: the alignment objective uses causal attention, while layer-wise refinement uses full attention. A key finding of our work is that text--vocal alignment can emerge within pure acoustic-token language modeling, without requiring a separate semantic token stage. In addition, we find that initializing the super-resolution model from the trained backbone significantly improves convergence and final quality.

	\subsection{Contributions}
	Our main contributions are summarized as follows:
	{\setlength{\leftmargini}{1.2em}
	\begin{itemize}
	\item \textbf{Deep acoustic token hierarchy.}
	We build a deep 64-layer RVQ acoustic token hierarchy, enabling a pure acoustic-token pathway toward higher fidelity.

	\item \textbf{Two-stage coarse-to-fine generation.}
	We propose a two-stage generation framework in a single acoustic token space: a backbone generates full-length coarse tokens, and a super-resolution model refines higher layers with full-length, time-parallel prediction, yielding a fixed 62-step inference process.

	\item \textbf{Hybrid-attention super-resolution training.}
	We introduce hybrid-attention training for super-resolution, which combines causal alignment learning and full-attention refinement within a single model, and show that it is critical for text--vocal alignment and lyric intelligibility.

	\item \textbf{Transferable coarse-to-fine priors.}
	We show that coarse-structure knowledge learned in the backbone is transferable to fine-grained refinement within the same acoustic token hierarchy, and leverage this with backbone-initialized super-resolution training to improve convergence and final quality.
	\end{itemize}
	}

\section{Related Work}

	\subsection{Audio Representations for Music Generation}

	Modern music generation is built on top of two broad types of audio representations: \emph{continuous latent spaces} and \emph{discrete token spaces}. Continuous latent approaches typically generate or refine music directly in latent space with diffusion-based models. These methods can achieve strong local rendering quality, but it is often more difficult for them to preserve long-range musical structure and other high-level content-related properties in full-length generation \citep{riffusion,stableaudio,diffrhythm}.

	Discrete representations form the other major line. Within this line, existing approaches can be roughly divided into \emph{semantic or semantic-like tokens} and \emph{acoustic codec tokens}. Semantic tokens are typically introduced to make long-range structure, conditioning, and high-level sequence modeling easier. This idea appears in hierarchical token-based systems such as AudioLM \citep{audiolm} and MusicLM \citep{musiclm}, and is also related to semantic-enriched codec designs such as MuCodec \citep{mucodec}, X-Codec \citep{xcodec} and HeartCodec \citep{heartmula}. Acoustic codec tokens, in contrast, are derived from neural codecs such as SoundStream \citep{soundstream}, EnCodec \citep{encodec}, HiFi-Codec \citep{hifi_codec}, DAC \citep{dac}, and WavTokenizer \citep{wavtokenizer}, and are optimized more directly for waveform reconstruction. They are therefore better suited for high-fidelity generation, but are also denser and harder to model directly.

	\subsection{Music Generation Pipelines}

    At the system level, recent music generation methods commonly adopt
    decomposed or hierarchical pipelines. A common pattern is to model
    high-level musical structure first, and then reconstruct or refine fine acoustic detail in a later stage. This appears in \emph{semantic-to-acoustic} pipelines such as AudioLM \citep{audiolm}, MusicLM \citep{musiclm}, YuE \citep{yue}, and MusiCoT \citep{musicot}; in \emph{acoustic-token hierarchies} such as Jukebox \citep{jukebox}; and in \emph{diffusion-based systems} such as SongBloom \citep{songbloom}, LeVo \citep{levo}, HeartMuLa \citep{heartmula}, and ACE-Step~1.5 \citep{acestep15}.

    More broadly, these systems reflect a common modeling question in music generation: how should high-level musical structure and fine-grained acoustic detail be organized across stages and representations? Existing work answers this in different ways, including semantic-to-acoustic factorization, acoustic coarse-to-fine hierarchies, and continuous latent diffusion. Our work revisits this question from a unified acoustic-token perspective, asking whether both can be progressively established within a single deep acoustic token hierarchy.

\section{Deep Acoustic Token Representation}
\label{sec:deep_acoustic_token_representation}

	\subsection{Residual Quantization for Audio}

	We represent music audio with a neural audio codec based on residual vector quantization (RVQ). Given an input waveform \(x\), the codec encoder maps it into a latent representation \(z\), which is then discretized by a stack of quantizers. Each quantization layer encodes the residual left by previous layers, and the decoder reconstructs the waveform from the sum of the quantized representations. As illustrated in Figure~\ref{fig:codec_overview}, this process converts waveform audio into a hierarchy of discrete acoustic tokens that can be used for downstream generation.

	Formally, the latent representation can be approximated as
	\[
	z \approx \sum_{n=0}^{N-1} q_n,
	\]
	where \(q_n\) denotes the discrete representation selected at the \(n\)-th quantization layer, and \(N\) is the total number of quantizers. In our setting, these quantized codes serve as the acoustic tokens for generation. Instead of generating waveforms directly, the models in the following sections operate on this hierarchy of discrete acoustic token layers, which we denote by \(q_0, q_1, \dots, q_{N-1}\).

	\subsection{Deep Acoustic Token Hierarchy}

\begin{figure}[t]
		\centering
		\includegraphics[width=\linewidth]{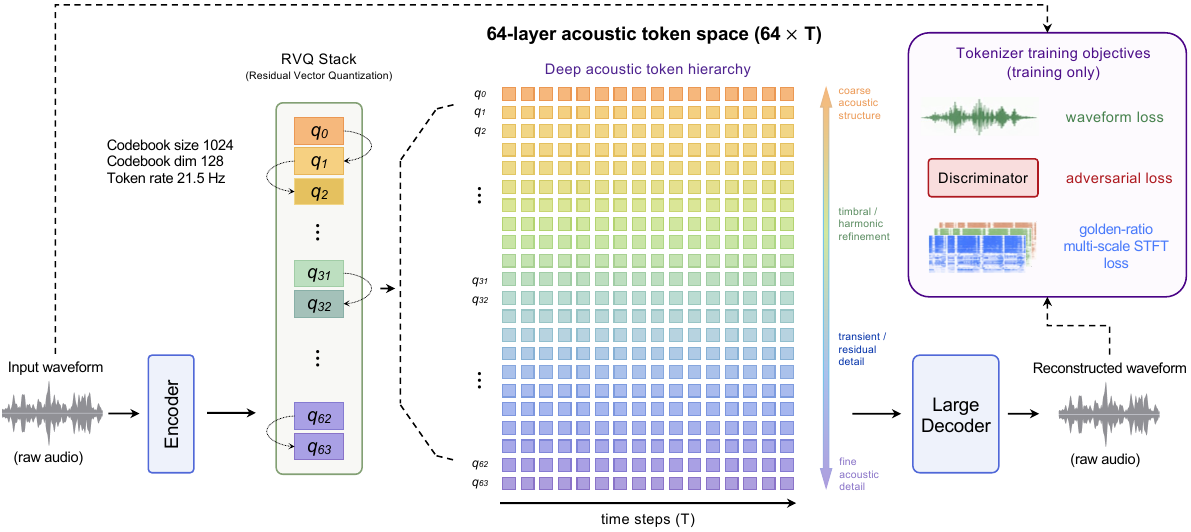}
		\caption{Overview of the neural audio codec and the resulting residual vector quantization hierarchy. Waveform audio is encoded into a stack of acoustic token layers, where lower layers capture coarse structure and deeper layers progressively encode finer residual detail.}
		\label{fig:codec_overview}
	\end{figure}

	A key property of residual quantization is that it naturally forms a hierarchical representation of audio. Lower quantization layers capture coarse acoustic structure, while deeper layers progressively encode finer residual details. As shown in Figure~\ref{fig:codec_overview}, the codec therefore does not only compress audio, but organizes it into a 2D acoustic token space over quantization layers and time. This coarse-to-fine structure is particularly important for music generation: lower layers tend to preserve broad acoustic content, while deeper layers contribute increasingly subtle information such as timbral refinement, harmonic detail, and transient structure.

	This hierarchical view also explains why shallow acoustic token spaces are often insufficient for high-fidelity music generation. A small number of quantization layers may be enough for rough reconstruction, but still leaves fine acoustic detail underrepresented. In practice, we choose 64 quantization layers as a reasonable operating point between reconstruction fidelity and downstream modeling complexity. In our preliminary experiments, shallower settings such as 16 layers still exhibit clearly perceptible distortion, whereas 64 layers already provide a qualitatively usable reconstruction regime. This choice is also supported by test-set reconstruction quality, where the 64-layer tokenizer reaches an average SDR of 9.116. Although even deeper hierarchies may further improve fidelity, they would also enlarge the token hierarchy and increase the difficulty of subsequent generation. We therefore use 64 layers as a practical balance in the present study. More importantly, this depth allows coarse and fine acoustic information to coexist in the same token space, providing the representational foundation for coarse-to-fine generation within a single acoustic hierarchy.

	\subsection{Challenges of Deep Quantization}

    The previous subsection motivates using a deep RVQ hierarchy, but such a hierarchy is not obtained simply by stacking more quantizers. Increasing the number of residual quantization layers makes the codec more expressive, since later layers are expected to capture increasingly fine acoustic residuals. At the same time, this makes optimization more delicate: the model must learn a stable division of labor across many residual layers while preserving both coarse structure and fine acoustic detail. In adversarial codec training, this also makes the balance between reconstruction losses and adversarial feedback harder to maintain. Without sufficient stability, deeper layers may become weakly used, redundant, or noisy rather than forming a useful coarse-to-fine representation.
    
    This issue directly affects downstream generation. The backbone model relies on the lowest layers as a stable coarse scaffold, while the super-resolution model relies on higher layers as predictable fine-grained refinement targets. We therefore treat stable training of the 64-layer tokenizer as a necessary part of the overall generation framework, and describe our training strategy in Section~\ref{sec:training_strategy}.

\section{Model Architecture}
    \label{sec:model_architecture}
	\subsection{Two-Stage Acoustic Token Generation}

	Given the deep acoustic token hierarchy introduced in Section~\ref{sec:deep_acoustic_token_representation}, the next question is how to model it with language models. A naive solution would be to flatten the entire hierarchy into a single 1D sequence and train one model to generate all quantization layers jointly. In our setting, however, this formulation is not practical. The hierarchy is deep, the resulting sequence becomes extremely long at full-track scale, and the model would need to handle both coarse generation and fine residual refinement within a single autoregressive process. This makes the optimization unnecessarily difficult and the decoding process prohibitively inefficient.

\begin{figure}[t]
		\centering
		\includegraphics[width=\linewidth]{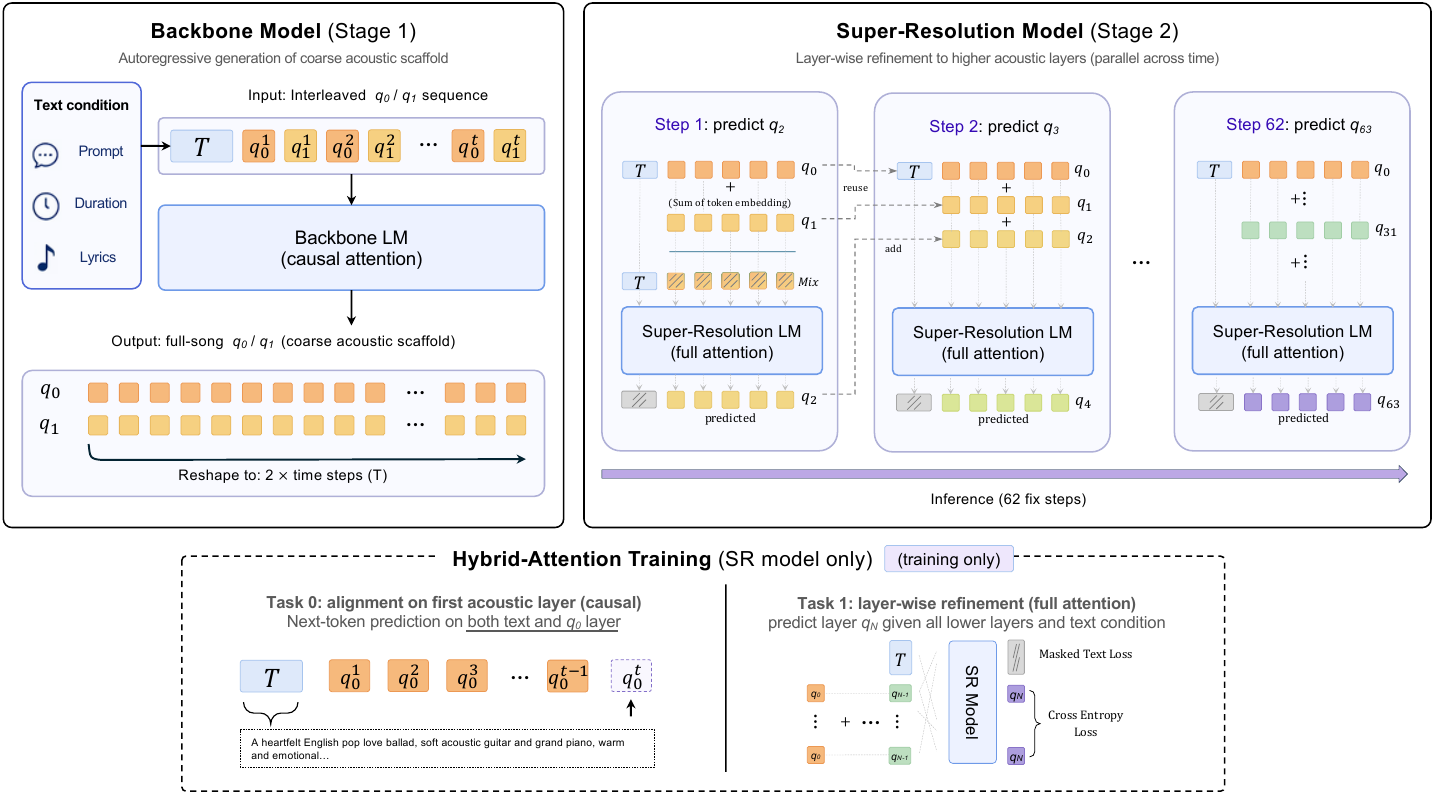}
		\caption{Overview of the proposed two-stage music generation framework. A backbone language model first generates coarse acoustic tokens for the full track, and a super-resolution model then refines the hierarchy layer by layer to recover higher-fidelity acoustic detail within the same token space.}
		\label{fig:model_overview}
	\end{figure}

	We therefore adopt a two-stage acoustic token generation framework. The first stage is a backbone model that autoregressively generates the low-level acoustic scaffold of the full track. The second stage is a super-resolution (SR) model that conditions on this scaffold and progressively predicts the remaining higher quantization layers. As shown in Figure~\ref{fig:model_overview}, this decomposition follows the coarse-to-fine structure of the acoustic hierarchy itself, while keeping the entire generation process within the same acoustic token space.

	\subsection{Backbone Model}

	The backbone model is responsible for generating the coarse acoustic scaffold of the generated music. Its input sequence consists of a text condition together with the low-level acoustic token stream. The text condition includes the information needed to control generation, such as the prompt, target duration, and lyrics. We tokenize this text input with the same text tokenizer used throughout the generation pipeline; implementation details are given in Section~\ref{sec:training_lms}.

	Concretely, the backbone operates on a sequence of the form
	\[
	[T] + [q_0^1, q_1^1, q_0^2, q_1^2, \dots, q_0^t, q_1^t, \dots, q_0^\tau, q_1^\tau],
	\]
	where \(T\) denotes the text condition and \((q_0^t, q_1^t)\) are the first two quantization layers at time step \(t\). The model is trained with standard next-token prediction under a causal attention mask. At inference time, it autoregressively generates the \(q_0\) and \(q_1\) layers for the full track, which then serve as the acoustic scaffold for the super-resolution stage. Training details are provided in Section~\ref{sec:training_lms}.

	\subsection{Super-Resolution Model}

    While the backbone establishes the coarse acoustic scaffold of the full track, high-fidelity generation still requires predicting the remaining higher quantization layers. We therefore introduce a super-resolution (SR) model that progressively refines the hierarchy conditioned on the lower layers. For each target layer \(q_N\), the SR model takes the text condition together with the sum of all previously available lower-layer quantized representations and predicts the full temporal sequence of the target layer:
    \[
    T + \sum_{i < N} q_i \rightarrow q_N.
    \]
    In this formulation, refinement is \emph{sequential across quantization layers} but \emph{parallel across time steps} within each layer. Since the backbone already provides the full-track \(q_0\) and \(q_1\) scaffold at inference time, the SR model starts from \(q_2\) and progressively predicts \(q_2, q_3, \dots, q_{63}\). Thus, the model operates directly at full-track scale, rather than refining local segments independently.
    
    This design also improves inference efficiency. After the backbone generates the coarse tokens for the full track, the SR model only needs to run one refinement step per remaining quantization layer. In our setting, this results in a fixed 62-step super-resolution process at full-track scale.

	\subsection{Hybrid-Attention Training for Super-Resolution}

    A natural starting point is to train the super-resolution model directly with the layer-wise refinement formulation described above, using full attention over the temporal dimension. At first glance, this seems sufficient: each refinement layer already has access to both the text condition and the lower-layer acoustic scaffold. However, in practice this formulation fails at a crucial point. While it can improve acoustic quality, it does not reliably produce vocals that are correctly aligned with the target lyrics. The problem is not that the model lacks access to text, but that the text is not temporally grounded in full-track refinement---that is, \textbf{the model does not reliably learn which part of the text should align with which local vocal realization over time.}

	This suggests a more specific view of alignment in pure acoustic-token generation. We do not treat alignment as requiring a separate semantic stage. Instead, we hypothesize that the lowest acoustic layer should serve as a \emph{text-grounded temporal scaffold}, on top of which finer acoustic detail can be progressively refined. Based on this view, we introduce an additional causal next-token prediction objective on the first acoustic layer. Concretely, given the text condition and the prefix of the first acoustic layer, the model predicts the next \(q_0\) token:
	\[
	[T,\; q_0^{<t}] \rightarrow q_0^t.
	\]
	The objective encourages the acoustic scaffold itself to become aligned with the temporal unfolding of the text. Empirically, we find that adding it immediately restores reliable lyric articulation and alignment, providing strong evidence that text--vocal alignment can emerge within pure acoustic-token language modeling when the learning objective is appropriately designed.

	This observation naturally leads to a hybrid-attention training scheme. The causal alignment objective is autoregressive and therefore uses a causal attention mask, whereas the layer-wise super-resolution objectives remain full-context and time-parallel, and therefore use full attention. For convenience, we refer to the causal alignment objective as Task~0, and to the layer-wise full-attention refinement objective as Task~1. Task~1 is instantiated across target layers: for a target layer \(q_N\), the model predicts \(q_N\) conditioned on the text and the sum of all lower layers. At inference time, only the full-attention refinement path is used; the role of Task~0 is purely to induce alignment during training. Detailed training formulations and ablations are presented in Sections~\ref{sec:training_strategy} and~\ref{sec:experiments}.

\section{Training Strategy}
\label{sec:training_strategy}

	\subsection{Training the Deep Acoustic Tokenizer}

	We train the acoustic tokenizer on an internal music dataset from the Central Conservatory of Music, containing about 18M tracks, or roughly 1.2M hours of audio. The corpus includes about 4.1M verified vocal songs paired with lyrics, about 4.8M verified instrumental tracks, and about 9.7M additional tracks without reliable vocal/instrumental tags. The dataset covers most major languages, and all audio is standardized to 44.1 kHz.

	Our tokenizer is a 64-layer RVQ neural codec operating at approximately 21.5 acoustic frames per second. Architecturally, it combines components inspired by both DAC and EnCodec: the encoder--decoder backbone follows a DAC-style design with a larger decoder for improved reconstruction quality, while the RVQ formulation and overall training pipeline follow EnCodec-style codec training. The encoder contains about 79M parameters, the decoder about 178M, and the quantizer about 8.4M.

	Training such a deep tokenizer is non-trivial. We optimize it with a combination of waveform reconstruction loss, adversarial loss, and multi-scale STFT loss. To make 64-layer quantization trainable in practice, we adopt two stabilizing design choices. First, we use a substantially stronger discriminator. At this depth, the codec generator becomes much more expressive, and a weak discriminator no longer provides a sufficiently strong adversarial signal. We therefore scale the discriminator to about 12M parameters with a base width of 128 filters, which is roughly 25$\times$ larger than the 0.48M discriminator used in EnCodec-style setups. We find that this stronger discriminator is important for stabilizing adversarial training at this scale. Second, following recent codec training practice~\citep{stablecodec}, we use a golden-ratio-inspired multi-scale STFT design with irregular FFT sizes [78, 126, 206, 334, 542, 876, 1418, 2296], with corresponding hop and window sizes following the same progression. Compared with more regular STFT grids, this design consistently yields better perceptual quality in our experiments. Overall, this recipe provides the best balance between fidelity and stability, and is crucial for preventing GAN collapse while training the 64-layer tokenizer.

	\subsection{Training the Language Models}
    \label{sec:training_lms}

	Both the backbone model and the super-resolution model are trained with the AdamW optimizer, a warmup of 2000 steps, and a global batch size of 256. Training is conducted on 8 H800 servers (64 GPUs) with gradient accumulation of 4. On the text side, both models use the Llama 3.1 8B tokenizer~\citep{llama3} with a vocabulary size of 128K. The shared transformer trunk contains 24 layers, hidden size 2048, 32 attention heads, 8 key-value heads, and FFN hidden size 5632, corresponding to about 1.2B parameters. Including embeddings, the final model sizes are about 1.6B for the backbone model and about 1.8B for the SR model, with the latter being larger due to its expanded vocabulary over text and acoustic token layers.

	For the backbone model, training samples are constructed by interleaving the first two acoustic layers along time, together with the text condition. The text condition includes prompts, duration control, and lyrics when available. We use a sequence length of 16384, which corresponds to an effective acoustic span of 8192 time steps after \(q_0/q_1\) interleaving. At a token rate of about 21.5 Hz, this covers roughly 381 seconds of music, or about 6 minutes, which is sufficient for most tracks in our dataset. The backbone is trained with standard next-token prediction under a causal mask.

	For the super-resolution model, training samples are organized as the task family introduced in Section~\ref{sec:model_architecture}, including the causal alignment task and the layer-wise refinement tasks. We use a sequence length of 8192, since the SR model predicts one target layer at a time while operating at full-track scale in parallel over time. In practice, the SR model is trained with a mixture of Task~0 and Task~1 instances under the corresponding hybrid-attention masks. We use a fixed empirically chosen task sampling mixture throughout SR training.

	\subsection{Backbone-Initialized Super-Resolution Training}

	Instead of training the super-resolution model entirely from scratch, we initialize it from a trained backbone checkpoint and then continue training on the SR objectives. This choice is motivated by the strong overlap between the two models: both operate in the same text-conditioned acoustic token space, and both rely on the low-level acoustic layers as the temporal scaffold for generation. The backbone learns to generate this scaffold, while the SR model must interpret it as the conditioning signal for fine-layer refinement.

	Conceptually, this initialization transfers coarse-structure and alignment-aware priors learned by the backbone into the super-resolution stage. In practice, we find that this significantly improves optimization: SR training converges faster and reaches a better final quality than training from scratch. We present quantitative comparisons in Section~\ref{sec:experiments}.

\section{Experiments}
\label{sec:experiments}

We evaluate Shao from two perspectives: large-scale human preference and ablations on the proposed training design. Since music generation is ultimately judged by human perception, we treat blind listening evaluation as the primary comparison protocol and first compare Shao against both commercial and open-source systems in a large-scale pairwise arena.

\subsection{Human Arena Evaluation}

We evaluate Shao using a large-scale blind pairwise listening arena that includes 8 music generation systems: 4 commercial models (Suno v5, Mureka v8, Suno v4.5, and MiniMax 2.5 Plus) and 4 open-source models (Shao, ACE-Step 1.5~\citep{acestep15}, HeartMuLa~\citep{heartmula}, and LeVo~\citep{levo}). The benchmark contains 10 prompts covering lyric-conditioned and text-guided music generation scenarios. For each comparison, raters listen to two anonymous samples and assign each sample an overall score on a 1--5 scale. After filtering out clearly unreliable raters---for example, those who completed the evaluation implausibly quickly or assigned nearly identical scores to all samples---the final arena contains 766 valid votes.

From these annotations, we report two complementary views of model performance. First, we compute the mean Overall Score, which reflects the absolute perceived quality of individual generations on a 1--5 scale. Second, we derive a Bradley--Terry (BT) score from the pairwise outcomes and convert it into an Elo-style ranking, which reflects relative strength in direct head-to-head comparison. The results are summarized in Table~\ref{tab:arena_main}.

Under BT-derived Elo, Shao ranks fourth overall and first among the open-source systems, slightly surpassing MiniMax 2.5 Plus in the current arena. Under mean Overall Score, Shao also ranks first among the open-source systems. In other words, Shao is not only the strongest open-source model in pairwise competition, but also the highest-rated open-source model under absolute human scoring.

The two rankings are closely related, but they are not expected to be identical. Mean Overall Score rewards models that consistently receive high standalone ratings across prompts and raters. BT-derived Elo, by contrast, emphasizes how often a model wins in direct pairwise comparisons, while also accounting for the strength of its opponents. As a result, a model may rank higher under absolute scoring by being consistently strong on average, while another may rank higher under BT by being particularly effective in head-to-head matchups, including narrow wins against strong competitors. We therefore view the two rankings as complementary rather than contradictory: one measures average subjective quality, while the other measures arena competitiveness.

For Shao, the encouraging point is that it remains strong under both views. This is especially noteworthy because Shao does not follow the dominant semantic-plus-decoder or diffusion-based formulation used by many contemporary systems. Instead, it is built on a pure acoustic-token route with a unified deep token hierarchy. Achieving the top open-source position under both absolute human scoring and pairwise arena ranking is therefore a meaningful result: it suggests that pure acoustic-token generation is not merely an interesting academic alternative, but already a practically competitive paradigm for high-quality music generation. Although the current arena is still moderate in scale, the present results indicate that this route is both viable today and highly promising for further scaling.

\begin{table*}[t]
	\centering
	\caption{\textbf{Main results on the blind pairwise human arena.}
We report both BT-derived Elo and mean Overall Score. ``Com.'' denotes commercial systems and ``Open'' denotes open-source systems. Shao ranks first among open-source systems under both views.}
	\label{tab:arena_main}
	\vspace{0.5em}
	\small
	\begin{tabular}{cclc}
		\toprule
		Rank & Src. & Model & BT-derived Elo \\
		\midrule
		1 & Com. & Mureka v8 & 1689.3 \\
		2 & Com. & Suno v5 & 1644.2 \\
		3 & Com. & Suno v4.5 & 1580.7 \\
		4 & Open & \textbf{Shao} & \textbf{1510.9} \\
		5 & Com. & MiniMax 2.5 Plus & 1509.2 \\
		6 & Open & ACE-Step 1.5 & 1470.9 \\
		7 & Open & HeartMuLa & 1421.8 \\
		8 & Open & LeVo & 1173.1 \\
		\bottomrule
	\end{tabular}
	\hspace{0.8em}
	\begin{tabular}{cclc}
		\toprule
		Rank & Src. & Model & Overall Score \\
		\midrule
		1 & Com. & Suno v5 & 3.9552 \\
		2 & Com. & Mureka v8 & 3.8583 \\
		3 & Com. & Suno v4.5 & 3.5907 \\
		4 & Com. & MiniMax 2.5 Plus & 3.5771 \\
		5 & Open & \textbf{Shao} & \textbf{3.3978} \\
		6 & Open & ACE-Step 1.5 & 3.3540 \\
		7 & Open & HeartMuLa & 3.2639 \\
		8 & Open & LeVo & 2.4515 \\
		\bottomrule
	\end{tabular}
\end{table*}

\subsection{Ablations on Alignment and Initialization}

We conduct targeted ablations to study two key components of our super-resolution training recipe: backbone initialization and Task~0. The first comparison evaluates the effect of backbone initialization when Task~0 is used. The second comparison evaluates the effect of Task~0 by comparing two backbone-initialized SR models at the same intermediate training stage. This allows us to examine both whether coarse-structure priors from the backbone improve final SR quality, and whether the causal alignment objective is necessary for reliable lyric intelligibility.

We report both subjective and objective metrics. For subjective evaluation, listeners rate audio quality and lyric intelligibility on a 1--5 scale. For objective evaluation, we use ASR-based lyric error rates as a proxy for intelligibility. We first transcribe generated vocals with Qwen3-ASR-1.7B~\citep{qwen3}, and then compare the recognized lyrics against the reference lyrics. Specifically, we report Chinese phoneme error rate (CN PER), Chinese character error rate (CN CER), and English word error rate (EN WER). ASR-based WER is not an absolute measure of human-perceived lyric correctness: generated singing voices may remain understandable to human listeners even when ASR makes transcription errors. We therefore interpret WER primarily as a relative objective proxy and report it together with human intelligibility ratings. The results are summarized in Table~\ref{tab:sr_ablation}.

First, backbone initialization improves the final SR model when Task~0 is used. As shown in Table~\ref{tab:sr_ablation}, the backbone-initialized model achieves higher subjective scores for both audio quality and lyric intelligibility, and also reduces bilingual lyric error rates across Chinese and English. In particular, CN PER decreases from 22.25\% to 16.67\%, and EN WER decreases from 26.96\% to 19.90\%. These improvements suggest that the backbone learns coarse-structure and text-conditioned acoustic priors that can be transferred to fine-layer refinement. In other words, the backbone does not only provide the first two acoustic layers at inference time; its learned representation also provides a better initialization point for training the SR model.

The effect of Task~0 is more directly tied to alignment. As discussed in Section~\ref{sec:model_architecture}, a layer-wise full-attention SR objective may improve acoustic quality, but it does not necessarily produce vocals that are correctly aligned with the target lyrics. The ablation results support this observation. Under fixed backbone initialization and the same intermediate training stage, the two models obtain similar audio-quality scores, but their lyric intelligibility differs dramatically. Adding Task~0 improves subjective lyric intelligibility from 1.34 to 2.70, while reducing CN PER from 78.67\% to 21.15\% and EN WER from 84.97\% to 25.15\%. This large gap indicates that access to text and lower-layer acoustic information is not sufficient by itself. Although lower-layer representations are still updated indirectly through higher-layer refinement losses, such indirect supervision does not reliably induce text--vocal alignment.

These ablations support the design motivation of hybrid-attention SR training. Backbone initialization provides transferable coarse-to-fine priors that improve the final refinement quality, while Task~0 supplies the explicit autoregressive grounding signal needed for temporal alignment. Together, they suggest that reliable lyric generation in a pure acoustic-token framework requires both effective transfer from the coarse acoustic scaffold and an explicit training objective that grounds the lowest acoustic layer in the temporal unfolding of the text.

\begin{table*}[t]
	\centering
	\caption{\textbf{Ablation results on alignment and initialization.}
Subjective audio quality and lyric intelligibility are rated on a 1--5 scale. CN PER, CN CER, and EN WER are ASR-based lyric error rates; lower is better.}
	\label{tab:sr_ablation}
	\vspace{0.5em}
	\begin{tabular}{llccccc}
		\toprule
		Ablation & Setting & Audio Quality $\uparrow$ & Lyric Intell. $\uparrow$ & CN PER $\downarrow$ & CN CER $\downarrow$ & EN WER $\downarrow$ \\
		\midrule
		Backbone Init.
		& w/o Init
		& 3.01 & 3.04 & 22.25 & 25.73 & 26.96 \\
		& w/ Init
		& \textbf{3.26} & \textbf{3.40} & \textbf{16.67} & \textbf{20.00} & \textbf{19.90} \\
		\midrule
		Task~0
		& w/o Task~0
		& 2.62 & 1.34 & 78.67 & 84.47 & 84.97 \\
		& w/ Task~0
		& \textbf{2.73} & \textbf{2.70} & \textbf{21.15} & \textbf{24.81} & \textbf{25.15} \\
		\bottomrule
	\end{tabular}
\end{table*}

\section{Discussion}
\label{sec:discussion}

Shao is unified at the level of acoustic token space, but the current system still uses a two-stage model design. We view this as a practical choice under the current compute budget, rather than a fundamental limitation of the paradigm. In preliminary experiments, we also explored a unified model that performs both coarse generation and super-resolution, but found that decoupling the two stages into separate models yielded better quality under the same parameter budget. Since both the backbone and the super-resolution model already operate in the same text-conditioned acoustic token hierarchy, their separation is mainly an optimization and scaling decision. A sufficiently large model with an appropriate training mixture could naturally absorb both coarse generation and fine-layer refinement into a single unified model. In this sense, the present two-stage system can be seen as an intermediate step toward a more integrated acoustic-token language model, where structure, alignment, and fidelity are learned within one model and one token hierarchy.

Another important direction is to further strengthen the acoustic tokenizer itself. Although the 64-layer RVQ tokenizer provides a practical balance between fidelity and modeling complexity, higher-quality generation will likely benefit from more robust deep quantization and stronger codec architectures. In particular, future tokenizers may require more powerful encoder--decoder backbones, more stable training objectives for very deep residual quantization, and better preservation of high-frequency details across vocals, instruments, and transients. Improving this acoustic foundation would directly benefit both stages of generation, and would also make fully unified acoustic-token modeling more feasible at larger scales.

\section{Conclusion}
\label{sec:conclusion}

We introduced Shao, a high-fidelity music generation system built around a deep acoustic-token hierarchy. Instead of relying on a separate semantic token stage or a diffusion-based decoder, Shao models music generation within a unified 64-layer RVQ acoustic token space. A backbone model first generates full-length coarse acoustic tokens, and a super-resolution model then refines the remaining layers at full-track scale with time-parallel prediction, resulting in a fixed 62-step refinement process.

A central finding of this work is that text--vocal alignment can emerge within pure acoustic-token language modeling, but only when the learning objective provides an explicit temporal grounding signal. Our hybrid-attention training scheme addresses this by combining a causal Task~0 objective for alignment with full-attention layer-wise refinement for acoustic detail. Ablations show that Task~0 is critical for lyric intelligibility, while backbone initialization further improves the final SR model across subjective and objective metrics.

Human arena evaluation shows that Shao ranks first among the evaluated open-source systems under both BT-derived Elo and mean Overall Score, and remains competitive with strong commercial baselines. These results suggest that pure acoustic-token generation is not merely an alternative formulation, but a practically competitive direction for high-quality music generation. More broadly, Shao points toward a simpler and more unified path for scaling music generation systems: a deep acoustic hierarchy, coarse-to-fine generation, and language-model-based learning throughout the pipeline.

\section*{Acknowledgments}

This work is supported by the Special Program of the National Natural Science Foundation of China (Grant No. T2341003) and the Major Program of the National Social Science Fund of China (Grant No. 21ZD19).

\bibliographystyle{unsrtnat}
\bibliography{references}

\end{document}